\documentclass[aps,prb,reprint, groupedaddress]{revtex4-2}
\usepackage[dvips]{graphicx}
\usepackage{amssymb}
\usepackage{bm}

\begin{document}

\title{Effect of retardation on the frequency and linewidth of plasma resonances in a two-dimensional disk of electron gas}

\author{I.V. Zagorodnev}
\email[]{zagorodnev@phystech.edu}
\affiliation{Kotelnikov Institute of Radioengineering and Electronics of Russian Academy of Sciences, Moscow, 125009 Russia}

\author{D.A. Rodionov}
\affiliation{Kotelnikov Institute of Radioengineering and Electronics of Russian Academy of Sciences, Moscow, 125009 Russia}

\author{A.A. Zabolotnykh}
\affiliation{Kotelnikov Institute of Radioengineering and Electronics of Russian Academy of Sciences, Moscow, 125009 Russia}

\date{\today}

\begin{abstract}
We theoretically analyze dominant plasma modes in a two-dimensional disk of electron gas by calculating the absorption of an incident electromagnetic wave. The problem is solved in a self-consistent approximation, taking into account electromagnetic retardation effects. We use the Drude model to describe the conductivity of the system. The absorption spectrum exhibits a series of peaks corresponding to the excitation of plasma waves. The position and linewidth of the peaks designating, respectively, the frequency and damping rate of the plasma modes. We estimate the influence of retardation effects on the frequency and linewidth of the fundamental (dipole) and axisymmetric (quadrupole) plasma modes both numerically and analytically. We find the net damping rate of the modes to be dependent on not only the sum of the radiative and collisional decays but also their intermixture, even for small retardation. We show that the net damping rate can be noticeably less than that determined by collisions alone.

\end{abstract}

\maketitle

\section{Introduction}

Plasma waves or plasmons in two-dimensional (2D) electron systems (ESs) were first discovered more than 40 years ago \cite{Grimes1976,Allen1977,Theis1977,Tsui1980}. Recognized as one of the basic and easily-excited collective oscillations, they have presently become a widely used platform for active fundamental and applied research, with considerable application potential in the fields of plasmonics, nanophotonics, and optoelectronics \cite{Knap2009,Grigorenko2012,Liu2015,Bandurin2019,Bylinkin2019,Fateev2019}. The interaction of plasmons with electromagnetic radiation is used ubiquitously to various studies in these areas.

Given an infinite homogeneous 2DES in vacuum environment, the dispersion of plasma waves was defined in \cite{Stern1967} as
\begin{equation} \label{eq:Stern}
	q^2=\frac{\omega^2}{c^2}+\left(\frac{\omega^2}{2\pi n e^2/m}\right)^2,
\end{equation}
where $n$ and $m$ are the electron concentration and effective mass, $q$ is the plasmon wave vector, and $c$ is the speed of light in vacuum (note that we use the CGS units throughout the presented analysis). The derivation of Eq.~(\ref{eq:Stern}) is based on the assumption of infinite electron relaxation time. The dispersion law that follows from Eq.~(\ref{eq:Stern}) is restricted to the region ''below'' the light cone $\omega=cq$. For this reason, the excitation of plasmons with electromagnetic radiation requires introduction of an inhomogeneity into the 2DES or external field. Therefore, metallic gratings \cite{Allen1977,Knap2009,Bylinkin2019,Polischuk2017,Maniyara2019}, near-field optical microscopy \cite{Grigorenko2012,Fei2012,Chen2012,Nikitin2016}, or samples with confined geometries, e.g. disks or strips \cite{Muravev2015,Muravev2016,Muravev2019b,Nikulin2020}, are utilized. The 2D disk, in particular, is one of the simplest configurations for both manufacturing and theoretical analysis --- fabricated without metallic electrodes, such a system is perfectly suitable for making a more direct comparison between the experimental data and theoretical calculations.

Plasmons in conductive 2D disks have been studied since 1985 \cite{Fetter1986,Glattli1985,Leavitt1986,Shikin1991,Ye1994}, and most recently, they have been intently discussed with regard to graphene structures \cite{Fang2013,Fang2014,Nikitin2016}. However, most of this work has been concerned with quasielectrostatic (quasistatic) or non-retarded regime, when the size of the sample is much smaller than the wavelength of the electromagnetic radiation. Retardation effects, on the other hand, significantly alter the properties of plasma oscillations, even in an infinite system \cite{Falko1989,Govorov1989,Levitov2020}. Thus, in a disk-shaped 2DES retardation affects plasmon spectra, drastically reduces the plasmon damping rate and considerably increases the quality factor~\cite{Kukushkin2003,Mikhailov2004,Muravev2017,Gusikhin2018,Muravev2019a}. At the same time, taking into account retardation effects greatly complicates the analytical treatment of plasma modes, which might explain the prevalent use of numerical approach in this case \cite{Balaban2009,Balaban2013,Mischenko2004,Forestiere2019}. 
In the presented paper we demonstrate the feasibility of qualitative and in some instances even quantitative analysis that accounts for the retardation effects.

In a disk-shaped 2DES, the eigen plasma modes are characterized by the radial number $n_r =  1,2,3,...$ and the orbital (angular) momentum number $l=0,\pm 1, \pm 2,...$ \cite{Glattli1985,Fetter1986,Balaban2009,Kukushkin2003}. The $l=0$ mode corresponds to the axisymmetric oscillations where the charges and currents move exclusively in the radial direction. This mode is also called dark (or breathing) mode since it has a zero dipole moment, and therefore, interacts rather weakly with electromagnetic radiation \cite{Schmidt2012,Muravev2017}. By contrast, in the $l = 1$ mode, the currents flow through the center as well as along the edge of the disk. The frequency of this mode is lower compared to the axisymmetric mode.

To estimate the frequency of plasma resonances in a finite-size sample, it is common to apply the phenomenological ''quantization rule'' of plasmon wave vector $q$ in dispersion law~(\ref{eq:Stern}). For example, according to the rule in a disk-shaped 2DES with the radius $R$, the fundamental mode ($l=1$, $n_r=1$) in the quasistatic regime is well-described by $q \approx 1.1/R$. This approximation is sufficiently accurate for calculating the plasmon frequency and even the damping rate when electromagnetic retardation effects are neglected. However, taking the retardation into consideration results in an additional contribution to the net plasmon damping rate due to electromagnetic radiation. This contribution is not described by the dispersion law~(\ref{eq:Stern}) as in derivation of it only non-radiative (localized near 2DES) modes were considered. Consequently, the overall damping rate of the plasmon resonances in finite-size 2DESs cannot be fully described by a simple quantization rule of the plasmon wave vector.

In this paper we analyze the frequency and damping rate of the plasma modes in a 2D disk for angular momentum $l=0$ and $l=1$ by calculating the absorption power of an electromagnetic wave. First, we transform the Maxwell's equations into an integro-differential equation for the current density. Then, we expand the unknown components of the current density in a Taylor-like series in a sense of the Galerkin method. Cutting the series we find the current density approximately. Finally, we determine the dependence of the frequency and linewidth of the absorption peaks on a retardation parameter. The absorption maxima indicate the excitation of plasma waves.

The two factors affecting the linewidth are the collisional (or dissipative) decay rate inversely proportional to the electron relaxation time and radiative decay rate related to the emission of electromagnetic radiation. Although most often the linewidth is assumed to represent purely additive effects of this two kinds of damping \cite{Mikhailov1996,Zoric2011,Andreev2014,Andreev2015}, we find that the linewidth is not merely the sum of these two decays. It contains additional mixing contributions. To the best of our knowledge, such intermixture of the plasmon dampings have been discussed for the first time. Furthermore, we obtain analytical approximations for the frequency and damping rate of the plasma modes with $l=0$ and $l=1$ excited in 2D disk.

In the following sections of the paper, we introduce the essential equations along with solution methods (Sec. II), expand on the specifics of the axisymmetric (Sec. III) and fundamental modes (Sec. IV), and close with discussion and conclusions (Sec. V).

\section{Key equations}

Consider a 2D electron-gas disk of radius $R$ in vacuum in the plane $z=0$. Let ${\bf r}=\left(x,y\right)$ be the radius vector in the disk plane. Following the approach in Ref.~\cite{Mikhailov2005}, we seek the system response to an incident external electric field ${\bf E}^{ext}\left({\bf r}\right)e^{-i\omega t}$ with oscillation frequency $\omega$. The total electric field ${\bf E}^{tot}\left({\bf r}\right)$ represents the superposition of the external field and the field induced by electron density in disk, ${\bf E}^{ind}\left({\bf r}\right)$. According to the theory of linear response, the current density in the disk becomes
\begin{equation} \label{eq:Current}
	{\bf j}\left({\bf r}\right) = \sigma\left(\omega\right) {\bf E}^{tot}\left({\bf r}\right) = \sigma\left(\omega\right) \left[ {\bf E}^{ext}\left({\bf r}\right) + {\bf E}^{ind}\left({\bf r}\right) \right].
\end{equation}
We consider Drude model for conductivity $\sigma\left(\omega\right) = ne^2\tau/m\left(1-i\omega\tau \right)$, where $n$, $m$, and $\tau$ are the 2D concentration, effective mass, and carriers relaxation time, respectively.

In fact, the Drude conductivity is governed by only two independent parameters, intrinsic to the system --- the collisional damping rate $\gamma = 1/\tau$ and $ne^2/m$, with the frequency $\omega$ being an extraneous parameter from the standpoint of the dynamical response. The internal properties can be varied nearly independently, even within a single sample, for example, by changing the temperature or carrier concentration~\cite{Kukushkin1989}. In the case of restricted systems, the size of the system becomes an additional parameter. However, it is convenient to introduce the following dimensionless parameters 
\begin{equation} \label{eq:Param}
	\widetilde{\gamma} = \frac{\gamma R}{c} = \frac{R}{c\tau}, \quad \widetilde{\Gamma} = \frac{2\pi n e^2R}{mc^2}. 
\end{equation}
It will be shown that these very parameters determine the characteristics of plasma waves in a disk. We refer to $\widetilde{\Gamma}$ as the retardation parameter. Notice, that these parameters (without $R/c$ factor) define the absorption line broadening in an infinite 2DES in the presence of magnetic field \cite{Chiu1976,Mikhailov2004}.

 Based on Eqs.~(\ref{eq:Param}), the conductivity can be rewritten in terms of the dimensionless parameters as
\begin{equation}  \label{eq:DrudeConductivity}
	\sigma\left(\widetilde{\omega}\right) = i\frac{c}{2\pi}\frac{\widetilde{\Gamma}}{\widetilde{\omega}+i\widetilde{\gamma}},
\end{equation}
where $\widetilde\omega = \omega R/c = 2\pi R/\lambda$ is the dimensionless frequency, which is equal to the ratio of the disk perimeter to the wavelength $\lambda$ of the external radiation. Here, the case of $\widetilde\omega \ll 1$ corresponds to the quasistatic regime. In the following analysis, we focus mainly on the dependence of plasmon characteristics on the retardation parameter $\widetilde\Gamma$ for any values of $\widetilde\omega$.

In practical sense, considering standard high-mobility GaAs/AlGaAs quantum wells with typical 2D electron concentration $n \sim 10^{11}$ cm$^{-2}$, for the radius of disk-shaped samples of up to $1.2$ cm, the retardation parameter $\widetilde\Gamma$ can reach the value of $10$~\cite{Gusikhin2018}, whereas the dimensionless relaxation rate $\widetilde\gamma$ is less than or on the order of unity.

Mathematical analysis of plasmons poses quite a challenge since the relationship between the current density and the induced electric field is nonlocal. For the self-consistent derivation of the electric field induced in a disk we first obtain the diagonalized system of scalar and vector potentials in the cylindrical coordinates $\left(r,\theta,z\right)$. Next, we apply the Hankel transformation to express the potentials through the current densities in the disk. After that, we take the inverse Hankel transform and apply the inverse diagonalization. Finally, we express the induced electric field, $\bm E^{ind}=(E^{ind}_r,E^{ind}_{\theta})^{T}$, through the potentials and connect it with the current $\bm j=(j_r,j_{\theta})^{T}$:
\begin{equation} \label{eq:InducedE}
	{\bf E}^{ind}(r) = i\frac{2\pi}{\omega}
	\left(\frac{\omega^2}{c^2}+ \hat{D} \right)
	\int\limits_{0}^{R}G_l(r,r'){\bf j}(r')r'dr',
\end{equation}
where
\begin{equation}
	\hat{D} = \left(\begin{array}{cc}
	\frac{1}{r}\frac{d}{d r}r\frac{d}{d r} - \frac{1}{r^2} \quad & il\frac{d}{d r}\frac{1}{r}\\
	\frac{il}{r^2}\frac{d}{d r}r \quad & -\frac{l^2}{r^2}
	\end{array}\right).
\end{equation}
Additional derivation details as well as the specific kind of kernel $G_l(r,r')$ are provided in the Appendix A. In equations above, $G_l(r,r')$ is an integral operator related to the Hankel transform. Once the dimensionless  coordinate $r/R$ is introduced, the electric field in the Eq.~(\ref{eq:InducedE}) is governed solely by $\widetilde{\omega}$ --- the dimensional frequency.

Now let us consider the behavior of the current density at the center and at the edge of the disk. The normal to the edge component of the current should vanish at the edge, i.e. $j_r(R) = 0$. At the center of the disk the behaviour is far more complex. Given the continuity equation
\begin{equation} \label{eq:continuity}
	-i\omega\rho (r) + \frac{d j_r(r)}{d r} + \frac{j_r(r)+ilj_\theta(r)}{r} = 0,
\end{equation}
the current density relation at the center of the disk becomes 
\begin{equation}\label{eq:CenterBC}
	j_r(0)+ilj_\theta(0) = 0
\end{equation} 
assuming there are no singularities in charge density and derivative of the current density. For the axisymmetric mode, it immediately leads to $j_r(0) = 0$. For $l\neq 0$ further Taylor series expansion of the current density about $r=0$ results in an extra condition $2j'_r(0) + i l j'_\theta(0) = 0$ (with the prime denoting the derivatives with respect to $r$), which ensures zero charge at the center of the disk.

To calculate the response, one should solve the Eqs.~(\ref{eq:Current}) and (\ref{eq:InducedE}) with above mentioned boundary conditions.
However, obtaining the exact solution to these equations is virtually unachievable. Therefore, to find an approximate solution, we expand the unknown vector-function ${\bf j}\left({\bf r}\right)$ in a complete set of basis functions, integrate the system over $r$, and then reduce it to a matrix equation on the expanding coefficients. After that, we truncate the matrix to retain only the dominant terms. Finally, by solving the resultant matrix equation, we calculate the expansion coefficients and, consequently, the desired current density, which determines the response of the system. The accuracy of this procedure can be assessed by the successive increase of the number of basis functions.

Although, any complete set of functions can be chosen, it is most appropriate to consider functions that are analytically integrable with the kernel (\ref{eq:DiskKernel}) of the integral equation (\ref{eq:InducedE}). Fortunately, any power function is a suitable choice since it can be integrated analytically with respect to the coordinates $r$, $r'$ at least with the inner kernel of the integral operator $G_l(r,r')$ (however, the integral over the parameter $p$ of the Hankel transform retains for some cases). Also, it is worth noting that due to the kernel properties, the current ${\bf j}({\bf r})$ has parity $(-1)^{l+1}$, as  discussed in the Appendix A. All the characteristics mentioned above constitute the key mathematical features that permit the exact calculation of most given intagrals, yelding analytical expressions for the current density and related plasma characteristics. 

Having determined the current density in the system, we calculate the absorption power
\begin{equation}\label{eq:AbsPower}
	P(\omega)=\int\limits_{0}^{R} \frac{1}{2}Re\left({\bf j^*}\cdot{{\bf E}^{tot}}\right)2\pi rdr=\frac{\pi\gamma}{2\Gamma}\int\limits_{0}^{R}\left|{\bf j}\right|^2rdr,
\end{equation}
which provides us with information about the position and width of the plasma resonances.

%%%%%%%%%%%%%%%%%%%%%%%%%%%%%%%%%%%%%%%%%%%%%%%%%%%%%%%%%%%%%%%%%%%%%%%%%%%%%%%%%

\begin{figure}
	\includegraphics[width=\linewidth]{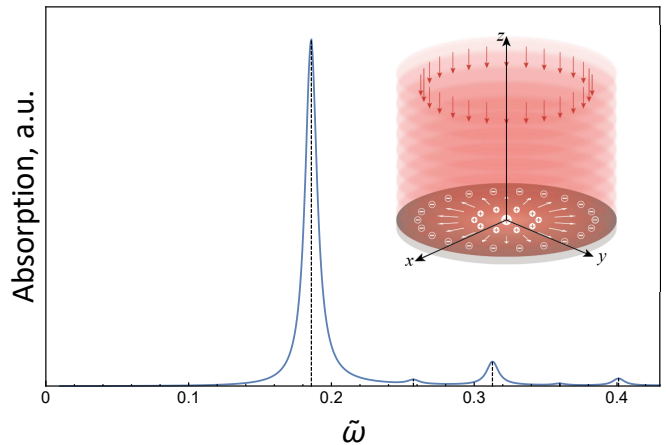}
	\caption{\label{fig:1} (Color online) Dependence of the absorption power (in arbitrary units) of the axisymmetric plasma mode on the dimensionless frequency of the incident radiation $\widetilde\omega = \omega R/c$, with $R$ and $c$ denoting the disk radius and the speed of light. Calculations results are carried out for the dimensionless collisional and retardation parameters $\widetilde\gamma = \widetilde\Gamma=0.01$, which are determined by the Eqs.~(\ref{eq:Param}), and ten terms of the current density expansion series in the Eq.~(\ref{eq:jrN}). Plotted data indicate the first five resonances with $n_r=1,2,...,5$. The inset schematically illustrates the excitation of the axisymmetric mode in the disk by external radiation.}
\end{figure}

\begin{figure}
	\includegraphics[width=\linewidth]{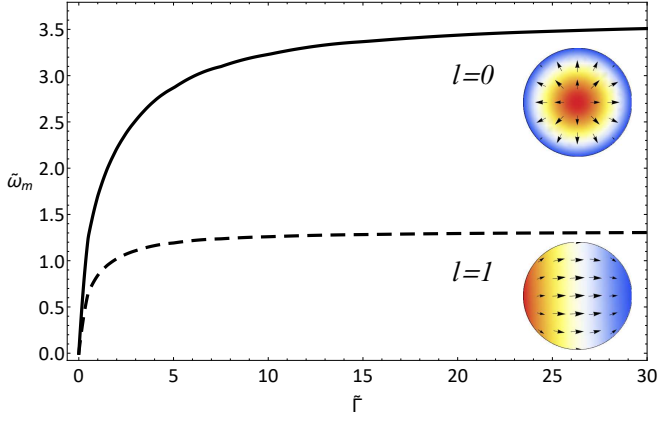}
	\caption{\label{fig:2} (Color online) Dependence of the dimensionless resonance frequency $\widetilde\omega_m$ on the retardation parameter $\widetilde\Gamma$ calculated for the plasma resonances of the radial number $n_r=1$ and the orbital numbers $l=0$ (solid curve) and $l=1$ (dashed curve). Calculations are carried out using ten basis functions for the $l=0$ mode and five basis functions for $l=1$ mode. Insets depict the charge and current distributions for the respective resonance modes.}
\end{figure}

\begin{figure}
	\includegraphics[width=\linewidth]{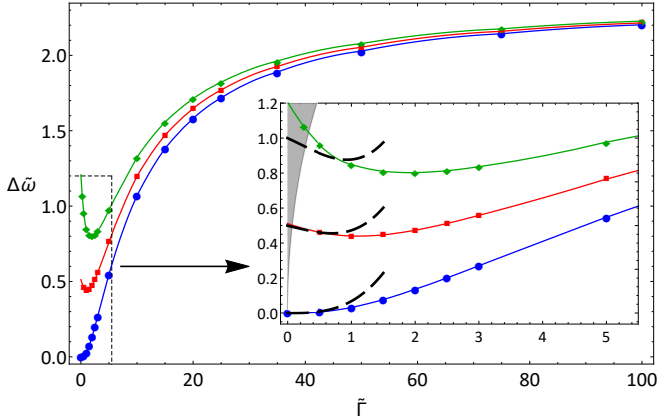}
	\caption{\label{fig:3} (Color online) Dependence of the dimensionless linewidth $\Delta\widetilde\omega = \Delta\omega R/c$ on the retardation parameter $\widetilde\Gamma$ calculated for the axisymmetric plasma resonance ($l=0$, $n_r=1$). Curves marked by (blue) circles, (red) squares and (green) diamonds correspond to the dimensionless collisional damping $\widetilde\gamma=0$, $\widetilde\gamma=0.5$ and $\widetilde\gamma=1$, respectively. Calculations are based on the first ten terms in the Eq.~(\ref{eq:jrN}). The inset is a close-up of the data region of small retardation, with dashed lines showing the approximations from the Eq.~(\ref{eq:linewidth1}). The grey shaded area designates the region with strong plasmon damping, $\Delta\widetilde\omega > \widetilde\omega_m$.
	}
\end{figure}

\begin{figure}
	\includegraphics[width=\linewidth]{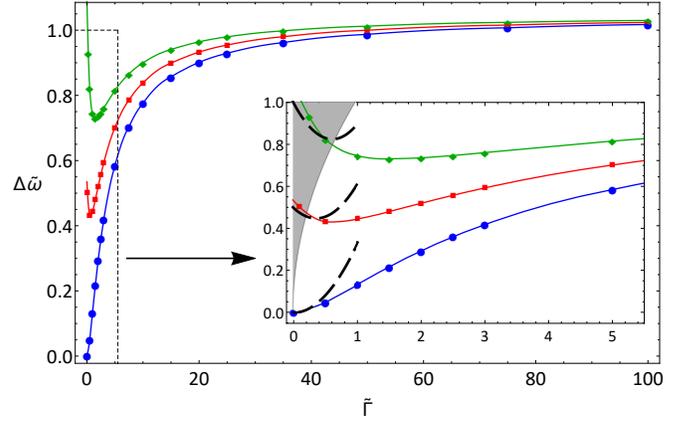}
	\caption{\label{fig:4} (Color online) Dependence of the dimensionless linewidth $\Delta\widetilde\omega$ on the retardation parameter $\widetilde\Gamma$ calculated for the fundamental resonance with $l=1$, $n_r=1$. Blue circle, red square and green diamond correspond to $\widetilde\gamma=0$, $\widetilde\gamma=0.5$ and $\widetilde\gamma=1$, respectively. The inset is a close-up of the data region of small retardation, with dashed lines indicating the approximations from the Eq.~(\ref{eq:linewidthl1}). The grey shaded area designates the region with strong plasmon damping.
	}
\end{figure}

%%%%%%%%%%%%%%%%%%%%%%%%%%%%%%%%%%%%%%%%%%%%%%%%%%%%%%%%%%%%%%%%%%%%%%%%%%%%%%%

\section{Axisymmetric mode ($l=0$)}

\subsection{Numerical solution}

For the axisymmetric mode the azimuthal components of the current and induced electric field are absent. Using the properties of the derivatives of the Bessel functions of the first order, which determine the kernel $G_0(r,r')$, the expression for the induced electric field can be reduced to
\begin{equation} \label{eq:ErInd}
	E_r^{ind}(r) = \frac{2\pi i}{\omega}\int\limits_0^\infty \beta J_1(pr)F(p) pdp,
\end{equation}
where $J_1(x)$ is the Bessel function of the first kind, $\beta = \sqrt{p^2-\omega^2/c^2}$, and the auxiliary function $F(p) = \int\limits_0^R j_r(r) J_1(pr) rdr$ is the Hankel transform of the current density $j_r(r)$. In the case $p<\omega/c$ the imaginary part of $\beta$ becomes negative, indicating the waves outgoing from the disk.

To excite the axisymmetric mode, we choose the external electric field of the form
\begin{equation} \label{eq:ErExt}
	{\bf E}^{ext} = \left(E_r^{ext}, E_\theta^{ext}\right)^T = E_0(1,0)^T
\end{equation}
The exact origin of this kind of field is irrelevant to the main subject of this paper. It may arise from a multipole expansion of a complex field. For example, as a result of a decomposition of excitation field in near-field scanning optical microscopy, or of an oscillating dipole (or antenna) in the vicinity of the disk.

For the radial component of the current we apply the series expansion 
\begin{equation} \label{eq:jrN}
	j_r(r)=\sum\limits_{n=1}^{N}\alpha_n \frac{r}{R}\left(1-\frac{r^2}{R^2}\right)^n,
\end{equation}
where $\alpha_n$ are unknown coefficients, and $N$ is the number of basis functions under consideration. The given basis set of functions can be orthogonolized, although we found it unnecessary. Comparing calculations results with those based on the orthonormal basis set (including all powers of $r$ as well as only odd powers), we observe no significant differences.

To obtain the equation for coefficient $\alpha_n$, we consecutively multiple the Eq.~(\ref{eq:Current}) by $k$-th basis function $\left(1-r^2/R^2\right)^kr/R$ for $k=1,2,...,N$ and then integrate the result with the weight function $r$ over the range $r \in [0,R]$. Here, all integrals are calculated analytically (including integral with respect to $p$). After that we arrive at a linear system of the unknown coefficients $\alpha_n$, which is easy to solve. Thus, having determined the current density, we calculate the absorption power spectrum, according to the Eq.~(\ref{eq:AbsPower}), and analyze it.

For the purpose of illustration, Fig.~\ref{fig:1} shows an example of the dependence of the absorption power on the frequency of external radiation, calculated for $\widetilde\gamma = \widetilde\Gamma=0.01$ in a quasistatic regime, with $\widetilde\omega\ll 1$ (and $\widetilde\Gamma \ll 1$), using the first ten basis functions. The data clearly indicates the first five resonances for $n_r=1,2,..,5$, with more pronounced excitation of the mode with odd rather than even values of $n_r$. Also, the ratio of the resonant frequencies to that of the lowest ($n_r=1$) resonance consistently comes to the value of $\sqrt{n_r}$ with high degree of accuracy. Thus, it is evident that the phenomenological quantization rule of the 2D wave vector $q\approx 3.5n_r/R$ well describes the position of the resonances, at least for small damping.

Below we focus only on the main resonance with $n_r=1$. From the dependence of the absorption power on the frequency we find the resonant frequency $\widetilde\omega_m$ as well as the linewidth $\Delta\widetilde\omega$ taken as the full width at half maximum. In general, they both depend on $\widetilde\gamma$ and $\widetilde\Gamma$. In practical case of particular interest, $\widetilde\omega_m \gg \widetilde\gamma$, when the resonant peak is well resolved (i.e. the width of the peak is less than its position), the dependence of the resonant frequency on the retardation parameter is unique. This dependence is shown in Fig.~\ref{fig:2}. Thus, at small frequencies $\widetilde\omega_m \approx 1.87 \sqrt{\widetilde\Gamma}$, which is consistent with the Eq. (\ref{eq:Stern}) and the quantization rule $q\approx 3.5/R$. For large $\widetilde\Gamma$, when the plasmon dispersion (\ref{eq:Stern}) approaches the dispersion of light, $\omega = cq$, the resonant frequency tends to the asymptote $\widetilde\omega_m \approx 3.5$, which likewise conforms to the given quantization rule.

The dependence of the linewidth on the retardation parameter is depicted in Fig.~\ref{fig:3}. In general, it depends on the ratio between the collisional and retardation parameters. In the case $\widetilde\gamma \ll \widetilde\Gamma$, the linewidth clearly grows large with increasing $\widetilde\Gamma$. However, given sufficiently high collisional damping rate the linewidth actually decrease at small retardation. In the following discussion, we focus on this particular case and find the approximation of the linewidth dependence on the retardation parameter.

\subsection{Approximate solution}

To begin with, we approximate the current distribution by a single basis function
\begin{equation}\label{eq:jr1}
	j_r (r) = \alpha_1 \left(1-\frac{r^2}{R^2} \right) \frac{r}{R},
\end{equation}
where $\alpha_1$ is the only unknown coefficient. Then, the Hankel transform of the current becomes: $F(p) = 2\alpha_1 J_3(pR)/p^2$. Next, we substitute Eqs. (\ref{eq:ErInd}), (\ref{eq:ErExt}), (\ref{eq:jr1}) into Eq.~(\ref{eq:Current}),  multiply the resultant equation by $(1-r^2/R^2) r^2/R$, and integrate the product over the radius. For this calculation, we introduce the dimensionless coordinate $r/R$. After some algebraic manipulations, we arrive at
\begin{equation} \label{eq:OneCoeff}
	\alpha_1 = \frac{i24c\widetilde\Gamma E_0}{15\pi  \left[\widetilde\omega + i\widetilde\gamma - \frac{24\widetilde\Gamma}{\widetilde\omega} f(\widetilde\omega) \right]},
\end{equation}
where
\begin{widetext}
\begin{equation}
	f\left(\widetilde\omega\right) = -\frac{16}{105\pi} + \frac{2H_3(2\widetilde\omega)}{\widetilde\omega^4} + \frac{H_4(2\widetilde\omega)}{\widetilde\omega^3} - i\frac{\widetilde\omega^5-8\widetilde\omega^3+48J_3(2\widetilde\omega)+24\widetilde\omega J_4(2\widetilde\omega)}{24\widetilde\omega^4}.
\end{equation}
\end{widetext}
Here $H_n(x)$ is the Struve function.

Since the absorption power is proportional to $|\alpha_1|^2$, the resonance is determined by the minimum of the denominator of $\alpha_1$. In the limit $\widetilde\omega \rightarrow 0$ we have
\begin{equation} \label{eq:expandf}
	f(\widetilde\omega) \rightarrow \frac{16}{35\pi} - \frac{64\widetilde\omega^2}{2835\pi} - \frac{256\widetilde\omega^4}{155925\pi} - \frac{i\widetilde\omega^5}{4320}+...
\end{equation}

Leaving only first three terms in the expression (\ref{eq:expandf}) and neglecting the collisional damping, we determine the position of the absorption maximum $\widetilde\omega_m$ for $\widetilde{\Gamma}\ll 1$:
\begin{equation} \label{eq:omega1}
	\widetilde\omega_m^2 \approx 3.492\widetilde{\Gamma}\left( 1 - 0.172\widetilde{\Gamma} - 0.01406 \widetilde{\Gamma}^2 \right).
\end{equation}

In the lowest order in $\widetilde\Gamma$ (quasistatic limit), this result approximates that obtained earlier, with better than 1\% accuracy \cite{Fetter1986}. In the next order, the given plasmon frequency is in excellent agreement with experimental data reported in Ref. \cite{Muravev2019a}. However, at high $\widetilde\Gamma$ the frequency approaches asymptotic value $\widetilde\omega_m \approx 3.65$, which slightly differs from that obtained numerically.

The linewidth of the absorption resonance, $\Delta\widetilde\omega$, is related to the imaginary part of the denominator of $\alpha_1$. Hence, using the last term in the Eq.~(\ref{eq:expandf}) at $\widetilde{\omega}=\widetilde{\omega}_m$ we obtain
\begin{eqnarray} 
	\label{eq:linewidth1}
	\Delta\widetilde\omega \approx \widetilde\gamma + 0.068 \widetilde\Gamma^3+	\qquad	\qquad \qquad\nonumber\\
	\widetilde\gamma \widetilde\Gamma\left(-0.172  -0.058 \widetilde\Gamma + 0.04\widetilde\Gamma^2 \right) 
\end{eqnarray}
Evidently, the linewidth is comprised of three contributions: the collisional damping $\widetilde\gamma$, the radiative damping $\propto\widetilde\Gamma^3$, and the third term including the intermixture of $\widetilde\gamma$ and $\widetilde\Gamma$. Therefore, the resultant approximation (\ref{eq:linewidth1}) clearly demonstrates that the linewidth is not merely the sum of the collisional and radiation decays. Note that for $\widetilde\Gamma\ll 1$, the linewidth in fact decreases with increasing $\widetilde\Gamma$. In the next subsection, we consider these contributing factors in detail to shed some light on the interplay of these parameters.

\subsection{Qualitative description of the linewidth}

In this part of the paper, we give physical explanation of established linewidth dependencies in connection with the properties of the disk plasma mode. Consider an external electric field exciting plasma oscillations at a resonant frequency defined in the Eq.~(\ref{eq:omega1}). Provided that energy losses over an oscillation period are small compared with the energy stored in the mode (i.e., far from the shaded region in Fig.~\ref{fig:3}), the (dimensional) linewidth $\Delta\omega$ can be determined as $\Delta\omega = P/W$, where $P$ is the power loss averaged over the oscillation period, and $W$ is the energy stored in the plasma mode --- electromagnetic energy and kinetic energy of the carriers. Let us next find $P$ and $W$ for the current density specified in the Eq.~(\ref{eq:jr1}).

The net power loss can be treated as a sum of the losses associated with the Joule heating (caused by carrier collisions), $P_J$, and electromagnetic radiation, $P_{rad}$. From the differential form of the Joule heating equation, we find
\begin{equation}
	P_J = \frac{Re\left(\sigma^{-1}\right)}{2}\int\left| j_r(r) \right|^2 2\pi r dr = \frac{\pi^2\alpha_1^2\widetilde\gamma}{20c\widetilde\Gamma}
\end{equation}

At the same time, from the continuity Eq.~(\ref{eq:continuity}), we determine the charge density at the resonance frequency: $\rho(r) = -i2\alpha_1(1-2(r/R)^2)/(\omega_m R)$. In this case, the electric and magnetic dipole moments are absent due to the symmetry of the charge and current distributions. At low frequency (small retardation), the radiation of the mode is defined by the electric quadrupole moment $Q = i\pi\alpha_1 R^3 diag\{1,1,-2\}/(6\omega_m)$, where $diag\{\}$ is a diagonal matrix. Therefore, the radiative loss attributed to quadrupole radiation is given by:
\begin{equation}
	P_{rad} = \frac{\left| \dddot{Q} \right|^2}{180c^5} = \frac{\pi^2\alpha_1^2R^6\omega_m^4}{1080 c^5} \propto \widetilde\Gamma^2.
\end{equation}

The total energy $W$ is the sum of the carrier kinetic energy, $W_k$, and the electromagnetic energy, $W_{em}$. As long as losses are small, $W$ can be calculated virtually at any point in time. However, it is somewhat easier to compute it at the moment of the peak current --- in the absence of the charges and (non-radiating) electric fields. Hence, the kinetic energy can be formulated as:
\begin{equation}
	W_k = \int n \frac{m{\bf v}^2(r)}{2} 2\pi r dr = \frac{\pi}{\Gamma c} \int j_r^2(r)2\pi r dr,
\end{equation}
where ${\bf v}(r) = {\bf j}(r)/(en)$. At the same exact moment, the electromagnetic energy is determined by the azimuthal component of the magnetic field, $H_\theta$. Thus, some algebraic manipulation yields
\begin{equation}
	W_{em} = \int dV \frac{\left| H_\theta \right|^2}{8\pi} = \frac{\pi^2}{c^2}\int\limits_{\frac{\omega}{c}}^\infty \frac{pdp}{\beta} F^2(p).
\end{equation}
In the given derivation process, we exclude the integration interval from $0$ to $\omega/c$ as it corresponds to the emission of radiation, which leads to energy losses.

Finally, we arrive at the equation for the linewidth as follows:
\begin{equation}\label{eq:damping1}
	\Delta\omega = \frac{P_J+P_{rad}}{W_k+W_{em}},
\end{equation}
Substituting the explicit expression for the current into Eq.~(\ref{eq:damping1}) and calculating all the integrals gives exactly the approximation (\ref{eq:linewidth1}). In the lowest order in $\widetilde\Gamma$, the electromagnetic energy radiation and radiative power loss can be neglected to yield $\Delta\omega = \gamma$. In the next order, the radiative power loss can still be ignored. However, the first term in the expansion of the electromagnetic energy in $\widetilde\Gamma$ is a constant since any radial current produces circular magnetic fields above and below the disk, which contribute to the electromagnetic energy. Consequently, the linewidth is still proportional to $\gamma$, though it decreases with increasing retardation parameter. In the succeeding orders in $\widetilde\Gamma$, the radiative power loss must also be taken into account as it contributes to the overall damping. However, this effect is not very prominent being partially counteracted by the growing denominator of $W_{em}$, as well as stabilizing resonant frequency.

As follows from the derivation above, the approximation (\ref{eq:linewidth1}) is valid for a rather narrow range of the retardation parameters: $\widetilde\Gamma \ll 1$ and  $\widetilde\gamma \ll \sqrt{\widetilde\Gamma}$. However, as can be seen from the dashed curves in the inset to Fig.~\ref{fig:3}, the qualitative dependence of the linewidth on the retardation parameter is described relatively properly beyond the limits (but only for small $\widetilde\Gamma$). Indeed, retardation appears to affect the collisional linewidth, $\widetilde\gamma$, more dramatically than it follows from the approximate Eq. (\ref{eq:linewidth1}).

So far, we have demonstrated that the linewidth of the absorption peak relates to the plasmon damping, which is shown to be more than mere sum of the collisional and radiative decay rates. As the denominator in Eq. (\ref{eq:damping1}) is found to depend on the retardation as well as radiative loss, it makes the overall dependence of the linewidth on the retardation parameter more complex, which can lead to substantial reduction in the total damping compared to the collisional decay rate $\gamma$.

\section{Fundamental mode ($l=1$)}

To excite the plasma mode with orbital number $l=1$, we consider a circularly polarized electromagnetic plane wave incident normally onto the system, with the electric field in the plane of the disk given by ${\bf E}^{ext} = E_0(1,i)^T$. In this case, we expand the current density in an orthonormal set of basis vector-functions ${\bf j}\left( {\bf r} \right) = \sum C_n {\bf j}_n\left({\bf r}\right)$, with the scalar product $<{\bf j}_m\left({\bf r}\right), {\bf j}_n\left({\bf r}\right)> = \int {\bf j}_m^*\left( {\bf r}\right) \cdot  {\bf j}_n\left( {\bf r} \right) dS = \delta_{mn}$ \cite{Bergman1980,Forestiere2016}, where the integral is taken over the disk area, and $C_n$ are the expansion coefficients. The explicit functions are provided in the Appendix B.

Repeating the procedure outlined in Sec. III A, we substitute the given current expansion into Eq.~(\ref{eq:Current}), consecutively multiply it by the basis functions and integrate over the disk area. Finally, solving the resultant system of equations for the coefficients $C_n$, we find the desired current density. Examples of the data calculated for the resonance with $n_r=1$ and its linewidth are shown in Fig.~\ref{fig:2} and Fig.~\ref{fig:4}, where the numerical calculations are carried out based on the first five basis functions.

To obtain analytical expressions for the frequency and linewidth of the fundamental resonance $n_r=1$, we use only the first basis functions from the set, similar to Sec. III. We find the integral with respect to $r$ and $r'$ analytically, whereas that over the variable $p$ we split into two parts --- one integral from $0$ to $\omega/c$ and the other from $\omega/c$ to $\infty$. The former is evaluated approximately by expanding it to the third order of $\omega/c$ while the latter is evaluated analytically first, and then expanded to the third order of $\omega/c$. 
Analyzing the obtained coefficient $C_1$, we arrive at the approximate estimate of the resonant frequency
\begin{equation}
	\label{eq:omega11}
	\widetilde{\omega}_m \approx 1.07\sqrt{\widetilde{\Gamma}}\left(1-0.293\widetilde{\Gamma}\right),
\end{equation}
and the linewidth
\begin{equation}\label{eq:linewidthl1}
	\Delta\widetilde{\omega}\approx\widetilde{\gamma}+ 0.333\widetilde{\Gamma}^2+ \widetilde{\gamma}\widetilde{\Gamma}\left(-0.586+0.150\widetilde{\Gamma}\right).
\end{equation}
The qualitative difference between the linewidth of the modes with $l=0$ and $l=1$ is that the latter possesses a non-zero electric dipole moment. Therefore, the radiative power of the mode is proportional to $\widetilde\Gamma$ and, as a result, in Eq. (\ref{eq:linewidthl1}) there appears $\widetilde\Gamma^2$ term. 

At large $\widetilde\Gamma$ we find that the frequency approaches asymptotic value $\widetilde\omega_m \approx 1.4$, while the linewidth approaches $\Delta\widetilde{\omega}\approx 1.17$.

\section{Discussion and conclusions}

To make the analysis above more complete, we take into account the dielectric permittivity of the surrounding medium, $\epsilon$, by making the following replacement of the key parameters: $\widetilde{\omega}_m\rightarrow\sqrt{\epsilon}\widetilde{\omega}_m$, $\Delta\widetilde{\omega}\rightarrow\sqrt{\epsilon}\Delta\widetilde{\omega}$ and $\widetilde{\gamma}\rightarrow\sqrt{\epsilon}\widetilde{\gamma}$. As a result, we find that the broadening of the linewidth associated with the collisions does not change, while the radiative broadening decreases with increasing $\epsilon$. It may additionally reduces the dacay rate compared to $\gamma$.

In summary, we have studied numerically and analytically the fundamental (dipole) and axisymmetric (quadrupole) plasma modes in a 2D disk of electron gas taking into account retardation effects. We find that the frequency and the linewidth of the resonances can be fully described by two dimensionless parameters: $\widetilde{\gamma}$ corresponding to the collisional damping rate, and the retardation parameter $\widetilde{\Gamma}$ defined by Eq.~(\ref{eq:Param}). We establish that for weak collisions, $\widetilde{\gamma}\ll \widetilde{\omega}$, the dimensionless frequency of plasma resonances, $\widetilde{\omega}$, is defined only by the retardation parameter $\widetilde{\Gamma}$, as indicated by Eqs.~(\ref{eq:omega1}) and (\ref{eq:omega11}). As for the resonance linewidth, we discover that it cannot be fully described by the sum of collisional ($\widetilde{\gamma}$) and radiative ($\propto\widetilde{\Gamma}^2$ for dipole and $\propto\widetilde{\Gamma}^3$ for quadrupole modes) damping rates. The reason for such a complicated behavior of the linewidth is that with increasing retardation parameter, the radiation decay and the energy stored in the mode both grow simultaneously. The competition of these two processes leads to the non-monotonous dependence of the linewidth on the retardation parameter, as well as the narrowing of the linewidth compared to collisional damping at small values of $\widetilde{\Gamma}$, see Figs.~\ref{fig:3} and \ref{fig:4}.

\begin{acknowledgments}
	This work was supported by the Russian Science Foundation (project no. 16-12-10411). We are grateful to V. M. Muravev, I. V. Kukushkin, and V. A. Volkov for valuable discussions.
\end{acknowledgments}

\appendix
\section{Induced electric field}

To determine the electric field induced by the current density in a disk, consider the corresponding electrostatic $\varphi\left({\bf r},z,t\right)$ and vector ${\bf A}\left({\bf r},z,t\right)$ potentials described by Maxwell's equations in the Cartesian coordinate system, in CGS units:
\begin{eqnarray}
	&& \Delta {\bf A}\left( {\bf r},z,t \right) - \frac{1}{c^2}\frac{\partial^2}{\partial t^2} {\bf A}\left( {\bf r},z,t \right) = -\frac{4\pi}{c}{\bf j}\left({\bf r},t\right)\delta\left(z\right), \\
	&& \Delta \varphi\left( {\bf r},z,t \right) - \frac{1}{c^2}\frac{\partial^2}{\partial t^2} \varphi \left( {\bf r},z,t \right) = -4\pi\rho\left({\bf r},t\right)\delta\left(z\right).
\end{eqnarray}
Here $c$ is the speed of light, $\rho\left( {\bf r},t \right)$ and ${\bf j}\left( {\bf r},t \right)$ are, respectively, the charge density and density of current, and $\Delta$ is the three-dimensional Laplace operator.

The above equations can be expressed in the Lorenz gauge as
\begin{equation} \label{eq:lorenzgauge}
	\frac{1}{c}\frac{\partial}{\partial t}\varphi\left( {\bf r},z,t \right) + \text{div} {\bf A}\left( {\bf r},z,t \right)= 0
\end{equation}

Given the cylindrical symmetry, the system can be characterized by the angular momentum $l$ and radial number $n_r$. Hence, using the cylindrical coordinates $\left(r,\theta,z\right)$, we can express the vector quantity under consideration in terms of their radial $A_r(r,\theta,z,t) = A_x\cos\theta+A_y\sin\theta$ and azimuthal $A_\theta(r,\theta,z,t) = -A_x\sin\theta+A_y\cos\theta$ components, with $A_z\left( {\bf r}, t \right) = 0$ --- as there are no current sources to contribute to the  $z$-component of the vector potential. Then, applying the Fourier transformation with respect to time (i.e., considering the solutions of the form $\exp\left(il\theta-i\omega t \right)$), we reformulate the Maxwell's equations as follows
\begin{equation}\label{eq:max_in_cyl}
	\left(\begin{array}{cc}
		\Box_l-\frac{1}{r^2} & -\frac{2il}{r^2}\\
		\frac{2il}{r^2} & \Box_l -\frac{1}{r^2}
	\end{array}\right) 
	\left(\begin{array}{cc}
		A_r(r,z)\\ A_\theta(r,z) \end{array}\right) =-\frac{4\pi}{c}\left(\begin{array}{cc} j_r(r)\\ j_\theta(r)
	\end{array}\right)\delta\left(z\right)
\end{equation}
where $\Box_l = \frac{\omega^2}{c^2}+\frac{\partial^2}{\partial r^2}+\frac{1}{r}\frac{\partial}{\partial r}+\frac{\partial^2}{\partial z^2}-\frac{l^2}{r^2}$ is the time and angle Fourier transform of the d'Alembert operator in the cylindrical coordinates. As a next step, the scalar potential $\varphi$ is expressed through the vector potential from the Lorenz gauge in (\ref{eq:lorenzgauge}). Thus, using the transformation
\begin{equation}
	\bm{A}(r,z) = S\bm{A}_{S}(r,z), \quad S=
	\left(\begin{array}{cc}
		i & -i\\
		1 & 1
	\end{array}\right).
\end{equation}
we diagonalize the system in (\ref{eq:max_in_cyl}) to obtain
\begin{equation}\label{eq:diagonalA}
	\left(\begin{array}{cc}
		\Box_{l+1} & 0\\
		0 & \Box_{l-1}
	\end{array}\right)\bm{A}_S(r,z)=-\frac{4\pi}{c}\bm{j}_S(r)\delta\left(z\right)
\end{equation}
Then, taking the Hankel transform of the result, we find the vector potential to be
\begin{equation}\label{eq:Ajconnect}
	\bm{A}(r,0)=\frac{2\pi}{c}\int\limits_{0}^{R}G_l(r,r')\bm{j}(r')r'dr',
\end{equation}
with the following kernel
\begin{widetext}
	\begin{equation} \label{eq:DiskKernel}
	G_l(r,r')
	= \left[\int_0^\infty\frac{pdp}{\beta}
	S\left(\begin{array}{cc}
	J_{l+1}(pr')J_{l+1}(pr) & 0\\
	0 & J_{l-1}(pr')J_{l-1}(pr)
	\end{array}\right)S^{-1}\right].
	\end{equation}
\end{widetext}
Here, at $p<\omega/c$ we choose the branch of the square root relation $\beta = \sqrt{p^2-\omega^2/c^2}$ with the negative imaginary part since it corresponds to the waves outgoing from the disk.

Finally, we determine the induced electric field
\begin{equation}\label{eq:Eind}
	\bm{E}^{ind}(r) = \bm{E}^{ind}(r,0) = i\frac{c}{\omega}\left[\text{grad}\,\text{div}+\frac{\omega^2}{c^2}\right]\bm{A}(r,0),
\end{equation}
where $z$-components of the gradient and divergence disappear. Substituting the Eq. (\ref{eq:Ajconnect}) into (\ref{eq:Eind}), we arrive at the Eq. (\ref{eq:InducedE}) discussed in Sec. II of the paper. 

Importantly, the kernel $G_{l}\left(r,r'\right)$ has parity $(-1)^{l+1}$ with respect to the (formal) transformation $r \rightarrow -r$ since its defining Bessel functions of order $l\pm 1$ have parity $(-1)^{l+1}$. As the parity is preserved by the differential operator grad div in Eq.~(\ref{eq:Eind}), the induced field, and therefore the current density, can be odd and even functions for $l=0$ and $l=1$, respectively.

\section{Basis functions for mode $l=1$}

As follows from the Appendix A, in the fundamental mode, the basis vector-functions must be even with respect to the radius $r$ for both components of the current density. For this reason, we choose a polynomial sequence initiated with a quadratic function in $r$. However, it is not uniquely defined by the boundary conditions at the edge and center of the disk, as well as normalization. To eliminate the ambiguity, we consider the quasistatic regime of ${\bf j} \propto \text{grad} \phi$, where $\phi$ is a scalar function (electric potential). Consequently, the components of the current density become interrelated as: $j_\theta = il\int j_r(r')dr'/r$. Applying this to the basis set and using the Gram-Schmidt orthonormalizing process, we arrive at the following functions:
\begin{widetext}
	\begin{eqnarray}
	&&\bm{j}_1=3 \sqrt{\frac{3}{14}}
	\left(
	\begin{array}{cc}
	1-\widetilde{r}^2, & i\left(1-\frac{\widetilde{r}^2}{3}\right)
	\end{array}
	\right)^T,\quad
	\bm{j}_2=11 \sqrt{\frac{5}{238}}
	\left(
	\begin{array}{cc}
	1-\frac{81 \widetilde{r}^2-70 \widetilde{r}^4}{11 }, & i\left(1-\frac{27 \widetilde{r}^2-14 \widetilde{r}^4}{11 }\right)
	\end{array}
	\right)^T,\nonumber\\
	&&\bm{j}_3=23 \sqrt{\frac{7}{1054}}
	\left(
	\begin{array}{cc}
	1-\frac{378 \widetilde{r}^2-950 \widetilde{r}^4+595 \widetilde{r}^6}{23 }, & i\left( 1-\frac{126 \widetilde{r}^2-190 \widetilde{r}^4+85 \widetilde{r}^6}{23}\right)
	\end{array}
	\right)^T,\nonumber\\
	&&\bm{j}_4=\frac{117 }{7 \sqrt{62}}
	\left(
	\begin{array}{cc}
	1-\frac{370 \widetilde{r}^2-1750 \widetilde{r}^4+2695 \widetilde{r}^6-1302 \widetilde{r}^8}{13 }, & i\left(1-\frac{370 \widetilde{r}^2-1050 \widetilde{r}^4+1155 \widetilde{r}^6-434 \widetilde{r}^8}{39}\right)
	\end{array}
	\right)^T,\nonumber\\
	&&\bm{j}_5=\frac{59 }{7}\sqrt{\frac{11}{142}}
	\left(
	\begin{array}{cc}
	1-\frac{2565\widetilde{r}^2-19250\widetilde{r}^4+51940\widetilde{r}^6-57834\widetilde{r}^8+22638\widetilde{r}^{10}}{59}, & i\left(1-\frac{855 \widetilde{r}^2-3850 \widetilde{r}^4+7420 \widetilde{r}^6-6426 \widetilde{r}^8+2058 \widetilde{r}^{10}}{59}\right)
	\end{array}
	\right)^T,
	\end{eqnarray}
\end{widetext}
where $\widetilde{r}=r/R$. These results are used in calculations discussed in Sec. IV of the paper.

\bibliography{main}

\end{document}